\documentclass[aps,amsfonts]{revtex4}    
     \usepackage{amsmath}    
     \usepackage{amsfonts}    
     \usepackage{amssymb}    
     \usepackage{graphicx}

     \newcommand{\beqa}{\begin{eqnarray}}
     \newcommand{\eeqa}{\end{eqnarray}}
     \newcommand{\beq}{\begin{equation}}
     \newcommand{\eeq}{\end{equation}}
    \newcommand{\ba}{\begin{array}}
     \newcommand{\ea}{\end{array}}    
     
\begin{document}    
     \title{Einstein-Podolsky-Rosen correlations between two
    uniformly 
     accelerated oscillators}   
     \author {Serge Massar}
     \email{smassar@ulb.ac.be}   
     \affiliation{Laboratoire d'Information Quantique and Centre for
    Quantum Information 
     and Communication,  C.P.  165/59, Universit\'{e} Libre de
    Bruxelles, Avenue F. D. Roosevelt 50, 1050 Bruxelles, Belgium}   
     \author{Philippe Spindel}   
     \email{spindel@umh.ac.be}
     \affiliation{M\'ecanique et Gravitation, Universit\'{e} de 
     Mons-Hainaut,Acad\'emie Wallonie-Bruxelles     
     20, Place du Parc, 7000 Mons,    Belgium}
     \begin{abstract}
     We consider the quantum correlations, {\sl i.e.} the
    entanglement, 
     between two systems uniformly accelerated with identical
    acceleration $a$ in opposite Rindler quadrants which have reached thermal
    equilibrium with the Unruh 
     heat bath. To this end we study an exactly soluble model
    consisting of two oscillators 
     coupled to a massless scalar field in 1+1 dimensions. We find
    that for some values of 
     the parameters the oscillators get entangled shortly after the
    moment of closest approach. 
     Because of boost invariance there are an infinite set of pairs
    of positions where the oscillators 
     are entangled. The maximal entanglement between the oscillators
    is found to be approximately 1.4 entanglement bits. 
     \end{abstract}    \maketitle    
     
     \section{Introduction}    The ground state of a quantum field is
    an extremely structured state.    This is true even if the field is
    non interacting. A first indication  that this is the case is that
    the propagator between two spatially separated points    never
    vanishes, no matter how far apart the points are. That is
    correlations in vacuum extend over an infinite range. More subtle is
    that these correlations are such that in vacuum two spatially
    separated regions are entangled. This was first exhibited in
    \cite{Unruh76} by decomposing Minkowski space into two Rindler
    wedges, and quantizing the field in each wedge. It is also very closely related to phenomena like black hole evaporation or particle creation induced by cosmological expansion\cite{BD,GO}.  
     It was later shown that the ground state exhibits quantum non
    locality: it is in principle possible to violate Bell inequalities in
    the vacuum, both for free fields\cite{W1,W2,W3} and for interacting
    fields\cite{W4}, and that this property applies to almost any quantum
    state\cite{W5,HC}.
     Recently B. Reznik considered a specific model in which two
    localized detectors are coupled to the vacuum in space like separated
    regions and showed that the detectors can get entangled \cite{R}.  In
    \cite{RB} it was shown that the correlations between the detectors
    could be used to violate a Bell inequality. These studies were later
    extended to the case of more than two localized detectors
    \cite{Retal}.    
     \par
     In the present work we pursue the study of how uniformly
    accelerated    detectors in opposite Rindler wedges get entangled.
    Contrary to previous work \cite{R} we shall study the case where the
    detectors are in equilibrium with the Unruh heat bath. Nevertheless,
    although both detectors have thermalised, we shall show that they get
    entangled for certain choices of parameters and for certain relative
    positions.
     \par
\begin{figure}[t]
     \includegraphics[scale=0.6]{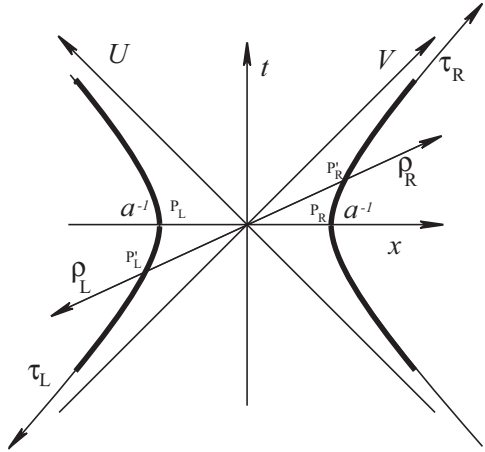}
     \caption{Minkowski and Rindler coordinates. The
    Minkowski time and space coordinates are $t$ and $x$; $V,U= t\pm x$ are the Minkowski
    light like coordinates; $\tau_R$, $\tau_L$ and $\rho_R$, $\rho_L$
    denote the Rindler time and space coordinates in the right and left
    Rindler quadrants respectively. Note that $\tau_R$ increases toward
    the future, whereas $\tau_L$ increases toward the past. The uniformly
    accelerated oscillators follow the trajectories $\rho_R  
    =a^{-1}$, and $\rho_L =
    a^{-1}$ which are indicated in the figure by bold curves.  Due to Lorentz invariance, 
    there is an infinite set of pairs of points for which the squared 
    invariant distance between the oscillators is minimal, such as 
    the pair $(P_{L},P_{R})$ or the pair 
    $(P^\prime_{L},P^\prime_{R})$ indicated on the figure. The two 
    oscillators only are entangled shortly after having reached 
    positions where $\Delta s^{2}$ is minimal. For instance if the 
    left oscillator is at $P_{L}$ (or $P^\prime_{L}$), the right 
    oscillator should be along its worldline, slightly after $P_{R}$ 
    (or $P^\prime_{L}$). }
     \label{fig:Traject}
     \end{figure}

     The model we    shall study consists of two oscillators
    uniformly accelerated with identical acceleration $a$ in opposite
    Rindler quadrants  coupled to a massless field in 1+1 dimensional
    Minkowski space time, see Fig. \ref{fig:Traject} for a depiction of
    the trajectories. This model is   exactly soluble, which allows us to
    study the case where the    oscillators are in  equilibrium with the
    Unruh heat    bath, and also to study the regime where the
    interaction between the    oscillator and field is strong, and 
    perturbation theory    is no longer valid.     We find
    (numerically) that, for certain values of the parameters, the two
    oscillators, both in thermal equilibrium with the Unruh heat bath,
    indeed get entangled. 
     An interesting aspect concerns the position at which the
    oscillators are maximally    entangled. Naively one would expect that
    this should occur when the  squared  invariant distance $\Delta s^{2}=\Delta x^2 - \Delta
    t^2$ between the oscillators    is minimal ({\sl i.e.} when
    the always spacelike separation between the two oscillators is minimal). But in fact, due to the
    dynamics of the oscillators which    absorb and emit quanta at a
    characteristic rate, the entanglement occurs only for a short period slightly after the moment
    of closest approach. Note also that because    of the boost
    invariance of the problem, there is  an infinite    set of
    pairs of points where the entanglement will be maximal.
     \par
     The model of an oscillator coupled to a massless field in 1+1
    dimensions has already been extensively studied, both in the context
    of the Unruh effect\cite{RSG,H,MPB,KK,K}, and with the aim of
    understanding    decoherence and thermalisation\cite{UZ,UW,ALZP}.
    Much of our analysis is based    on this earlier work. However during
    the present investigation we    encountered a problem which had
    apparently not been noticed before,    namely that the momentum of
    the uniformly accelerated oscillator has    infinite fluctuations due
    to an infrared divergence. We show how    these divergences can be
    controlled.
     \par
     Note that because the system we consider consists of two
    oscillators,    we must use the tools which have developed for the
    study of    entanglement of continuous variable systems, first 
    considered in the seminal paper of Einstein, Podolsky and Rosen 
    \cite{EPR}.  The    relevant tools will be reviewed below.    
     
     \section{The model}   
     We parametrize 1+1 dimensional Minkowski space as    
     \beqa    &t=\rho_R \sinh a \,\tau_R \quad , \quad x=\rho_R \cosh 
     a\,\tau_R \quad    (|t|< x)&  \quad ,  \nonumber\\    &t=-\rho_L \sinh a
    \,\tau_L \quad , \quad x=-\rho_L \cosh a\, \tau_L \quad    
    (x<-|t|)&\quad .
    \eeqa    
    Let us note that $\tau_{R}$ and $-\tau_{L}$ measure the proper 
    time along the trajectories ($ \rho_R = \rho_L = a^{-1}$) of the two oscilators.
     Later on we shall make use of the null coordinates    \beqa
    V&=&t+x=\left\{\ba{ll}     \ a^{-1}\exp a v_{R}&\quad (|t|< \ x)\\
    -a^{-1}\exp a v_{L}&\quad (x<-|t|)    \ea \right .\qquad ,\nonumber\\
    U&=&t-x=\left\{\ba{ll}    -a^{-1}\exp a u_{R}&\quad (|t|< x)\\    \
    a^{-1}\exp a u_{L}&\quad (x<-|t|)    \ea \right .\qquad .\nonumber\\    \eeqa

     Note that with these definitions $\tau_R$, $v_R$ and $u_L$
    increase toward the future, whereas $\tau_L$, $v_L$ and $u_R$
    increase toward the past.
     \par  
     The problem of a single uniformly accelerated oscillator coupled
    to a massless field in $1+1$ dimensions has been extensively
    studied\cite{RSG,H,MPB,KK,K}. Here we generalise it to the case of two
     oscillators  uniformly accelerated
    with  the same  acceleration $a$ in opposite Rindler quadrants. The
    action describing this system is   
     \beqa    S&=& \int dx dt \frac{1}{2} [ (\partial_t \phi)^2 -
    (\partial_x    \phi)^2] \nonumber\\    
     & &+ \int d\rho_R d\tau_R \delta (\rho_R - a^{-1})    \left[
    \frac{m}{2} \left( \frac{dq_R}{d\tau_R}\right)^2    -\frac{m}{2}
    \omega^2 q_R^2    
     + \epsilon \frac{dq_R}{d\tau_R} \phi (\tau_R, a^{-1})\right]
    \nonumber\\    
     & &- \int d\rho_L d\tau_L \delta (\rho_L - a^{-1})    \left[
    \frac{m}{2} \left( \frac{dq_L}{d\tau_L}\right)^2    -\frac{m}{2}
    \omega^2 q_L^2    - \epsilon \frac{dq_L}{d\tau_L} \phi (\tau_L,
    a^{-1})\right]    \eeqa   
     where $q_R$ and $q_L$ are the internal coordinates of the
    oscillators; and $m,  \omega$, $\epsilon$ are    their mass,
    oscillation frequency, and coupling to the field.    The momentum
    conjugate to the oscillator coordinates are   
     \beqa\label{pR}    p_{R}&=& m \frac{dq_{R}}{d\tau_{R}}+ \epsilon
    \phi (\tau_{R},    a^{-1})\nonumber\qquad  ,\\\label{pL}    p_{L}&=& - m
    \frac{dq_{L}}{d\tau_{L}}+ \epsilon \phi (\tau_{L}, a^{-1})   
    \qquad  .
    \eeqa   
     The  equations of motion are     
     \beqa    \partial_t^2\phi - \partial_x^2 \phi &=& \epsilon
    \frac{dq_R}{d\tau_R}    \delta (\rho_R - a^{-1})    + \epsilon
    \frac{dq_L}{d\tau_L} \delta (\rho_L - a^{-1})  \qquad ,  \nonumber\\
     m \frac{d^2q_R}{d\tau_R^2}+ m  \omega^2 q_R^2 &=& -\epsilon
    \frac{d \phi    (\tau_R , a^{-1})}{d\tau_R}\qquad ,\nonumber\\
     m \frac{d^2q_L}{d\tau_L^2}+ m  \omega^2 q_L^2 &=& + \epsilon
    \frac{d \phi    (\tau_L , a^{-1})}{d\tau_L} \qquad  .   \eeqa    
     The first of these equations may be integrated to yield    
     \beq
    \phi(x,t) = \phi^0(x,t)+ \frac{\epsilon}{2}q_R(\tau_{Rret}) +
    \frac{\epsilon}{2}q_L(\tau_{Lret})    
    \eeq     
    where $\tau_{R,L ret}$
    iare the values of $\tau_{R,L}$ at the    intersection of the past light
    cone from $(x,t)$ with the right and left    
    trajectories\footnote{When the intersections exist.} and where
    $\phi^0(x,t)$ is the free field operator,    solution of
    $\partial_t^2\phi^0 - \partial_x^2 \phi^0  =0$.    We can reexpress
    the free field solution as    
     \beq    \phi^0(x,t) = \mathop{\phi^0}\limits_{{\leftarrow}}(t+x)
    + 
     \mathop{\phi^0}\limits_{{\rightarrow}}(t-x) \qquad  .    \eeq    
     Inserting the solution for $\phi$ into the equation for
    $q_{R,L}$    yields the equations
     \beqa    m \frac{d^2q_R}{d\tau_R^2}+    \frac{\epsilon^2}{2}
    \frac{dq_R}{d\tau_R}     +m  \omega^2 q_R &=& -\epsilon \frac{d
    \phi^0 (\tau_R ,    a^{-1})}{d\tau_R}\qquad  ,\nonumber\\  
     m \frac{d^2q_L}{d\tau_L^2} -   \frac{\epsilon^2}{2}
    \frac{dq_L}{d\tau_L} + m  \omega^2 q_L &=& + \epsilon    \frac{d
    \phi^0 (\tau_L , a^{-1})}{d\tau_L}  \qquad  .  \eeqa   
     The equation for $q_R$ can be integrated to yield    \beq
    q_R(\tau_R)= q_0(\tau_R) - \int_{-\infty}^{\tau_R} d \tau_R'
    \chi(\tau_R-\tau_R')e \frac{d 
    \phi^0(\tau_R',a^{-1})}{d\tau_R'}\qquad  .
    \eeq    
     A similar solution obtains for $q_L$. Here $\chi$ is the
    retarded propagator    of $q$, and     $q_0$ is solution of
    $m\ddot{q_0}+\epsilon^{2} \dot{q_0} /2 + m  \omega^2 q_0=0$. The    free solution
    $q_0$ is exponentially damped. Henceforth we will    neglect it. This
    corresponds to supposie that the oscillators were    set into
    acceleration sufficiently far in the past and have reached thermal
    equilibrium with the Unruh heat bath.   
     \par 
     Taking  Fourier transforms, we can reexpress the solutions
    for the    oscillators as  
     \beqa\label{qR}    q_R(\tau_R)&=& i \frac{\epsilon}{m} \int d
    \lambda e^{-i\lambda \tau_R}    \lambda \chi_\lambda
    \phi^0_R(\lambda)\qquad ,\nonumber\\   
     \label{qL}    q_L(\tau_L)&=& i \frac{\epsilon}{m} \int d \lambda
    e^{+ i\lambda \tau_L}    \lambda \chi_\lambda 
    \phi^0_L(\lambda)\qquad ,
    \nonumber\\ 
     \chi_\lambda &=& \frac{-1}{ \lambda^2 -   \omega^2 + i
    \frac{\epsilon^2}{2    m} \lambda }\qquad ,\nonumber\\ 
     \phi^0_R(\lambda)&=& \int \frac{d \tau_R}{2 \pi} e^{+i \lambda
    \tau_R}    \phi^0(\tau_R,a^{-1}),\nonumber\\   
     \phi^0_L(\lambda)&=& \int \frac{d \tau_L}{2 \pi} e^{-i \lambda
    \tau_L} \phi^0(\tau_L,a^{-1}) \qquad  .   \eeqa
     \par
     Since the equations of motion are linear, these expressions
    represent both the solutions to the classical and Heisenberg
    equations of motion. In what follows we will take the initial state
    of the field $\phi$ to be the vacuum state. The initial state of the
    oscillator is then irrelevant, since as we noted above, its state
    depends only on the state of the field. 
     \par
     A key point in our analysis is that the oscillators are then in a
    Gaussian state. Indeed
     the initial state of the field, the vacuum, is a Gaussian state.
    And the internal coordinates $q_{L,R}$ and conjugate momenta
    $p_{L,R}$ of the oscillators depend linearly on the field operator.
    Therefore the oscillators are also in a Gaussian state. This means
    that the reduced density matrix of the oscillators, obtained by
    tracing over the field degrees of freedom, is entirely characterised
    by the expectation values of the first and second moments of the oscillator
    variables. In particular, as we will review in the next section, 
    these moments completely characterise the entanglement
    between the two oscillators.   
     \par
     It is immediate to obtain that the canonical variables have
    vanishing expectation value    $$    \langle q_R \rangle = \langle
    q_L \rangle = \langle p_R \rangle =    \langle p_L \rangle =0\ .$$    
     It is a more complicated task to compute the
    covariance matrix:    \beq  \label{covmat}  \sigma (\tau_R , \tau_L) = \left(
    \begin{array}{cccc}    \langle q_R (\tau_R) q_R(\tau_R)\rangle &
    \langle \{ q_R (\tau_R) , p_R(\tau_R) \} \rangle /2 &    \langle q_R
    (\tau_R) q_L(\tau_L)\rangle &    \langle q_R (\tau_R)
    p_L(\tau_L)\rangle \\    \langle \{ q_R (\tau_R), p_R(\tau_R) \}
    \rangle /2 &    \langle p_R (\tau_R) p_R(\tau_R) \rangle &    \langle
    p_R (\tau_R) q_L(\tau_L)\rangle &    \langle p_R (\tau_R)
    p_L(\tau_L)\rangle \\    \langle q_L (\tau_L) q_R(\tau_R)\rangle &
    \langle q_L (\tau_L) p_R(\tau_R) \rangle &    \langle q_L (\tau_L)
    q_L(\tau_L)\rangle &    \langle \{ q_L (\tau_L) ,p_L(\tau_L) \}
    \rangle /2 \\    \langle p_L (\tau_L) q_R(\tau_R)\rangle &    \langle
    p_L (\tau_L) p_R(\tau_R) \rangle &    \langle \{ q_L (\tau_L)
    ,p_L(\tau_L) \} \rangle /2	&    \langle p_L (\tau_L)
    p_L(\tau_L)\rangle     \end{array}    \right)    \eeq    where $\{
    \cdot , \cdot \}$ is the anticommutator. The covariance    matrix
    depends on the positions  $\tau_R$ , $\tau_L$ of the two
    oscillators. The expectation values along a single trajectory, such
    as $\langle q_R (\tau_R) q_R(\tau_R)\rangle$, are independent of the
    position along the trajectory, since we have supposed that the
    oscillators have reached a stationary state. The expectation values
    between    operators on opposite trajectories, such as $\langle q_R
    (\tau_R)    q_L(\tau_L)\rangle $ depend only on $T=\tau_R - \tau_L$,
    since by boost    invariance it depends only on the invariant
    distance between the two    oscillators : $\Delta 
    s^{2}=4\,a^{-2}\cosh a\,T/2$. Thus $\sigma$ is a function
    only of $T$.    The detailed calculation of $\sigma$ will be carried
    out in section    \ref{correlations}   and in the appendix.
     
     \section{Entanglement in continuous variable systems}  
     
     Entanglement in continuous variable systems has been extensively
    studied, see for instance \cite{Simon,GKLC,Ad}. We summarize here the
    results we will need in the remainder of the article.
     
     Consider two oscillators whose phase space variables $(q_R,p_R)$
    and $(q_L,p_L)$ obey the canonical commutation relations. It is
    convenient to group the phase space variables as 
     $\vec \xi = (q_R,p_R,q_L,p_L)$. We can write the canonical
    commutation relations as
     $$[ \xi_l , \xi_m ] = i {\pmb \Omega}_{lm}$$ where ${\pmb
    \Omega}$ is the symplectic matrix                    \beq    {\pmb
    \Omega}= \left( \ba{cccc}   0&1&0&0\\   -1&0&0&0\\   0&0&0&1\\
    0&0&-1&0    \ea \right)\qquad .    \eeq 
     
     For any quantum state $\rho$ of the two oscillators, we can
    compute the first and second moments of its phase space variables
     \beqa < \xi _l > &=& Tr \rho \xi_l\\
     {\pmb    \sigma}_{lm} &=& \frac 12 Tr \rho \{ \xi_l - < \xi _l >
    , \xi_m - < \xi _m > \}
     \eeqa
     where $\{ , \}$ is the anticommutator.
     The covariance matrix of the oscillators is a real, symmetric,
    positive matrix, satisfying the constraint (which follows from
    positivity of the Hilbert Schmidt norm)
     \beq \label{pos}
     {\pmb    \sigma} + i \frac 12 {\pmb \Omega} \geq 0\ .\eeq
     In general the first and second moments are only a partial
    characterisation of the quantum state $\rho$. But in the particular
    case where $\rho$ is Gaussian, they completely characterise the state.
     \par
     The correlation matrix ${\pmb    \sigma}$ allows one to study
    the entanglement of $\rho$. Denote by 
     $$\Lambda_{PT} = \left( \ba{cccc}   1&0&0&0\\   0&1&0&0\\
    0&0&1&0\\   0&0&0&-1    \ea \right)$$ the matrix which realises the
    partial transpose. A necessary condition for entanglement of the two
    oscillators is
     \beq
     \Lambda_{PT} {\pmb    \sigma} \Lambda_{PT}+ i \frac 12 {\pmb
    \Omega} \leq 0\label{condent}\ .\eeq
     This condition is also sufficient if the oscillators are in a
    Gaussian state.
     It is convenient to rewrite this entanglement conditions as
    follows. Express the covariance matrix (\ref{covmat}) as a bloc matrix   
     \beq  V=\left(\ba{cc} {\bf A}&{\bf C}\\   {\bf C}^{ T} &{\bf B}
    \ea \right) \qquad ,  \eeq 
     then eq. (\ref{condent}) is equivalent (when ${\bf A}={\bf B}$)
    to    \beq\label{Wil}   {\bf W}=\frac 12( \det[{\bf A}]-\det[{\bf
    C}])-\det[{\pmb    \sigma}]-\frac 1{16} < 0 \qquad.   \eeq  
     \par
     Below we will use the logarithmic negativity $E_N$ as quantitative measure of entanglement. It is defined as
       \beq E_N = \max[0, - \ln 2 \eta^-] \label{EN}\eeq
       where $\eta^-$ is the smallest
   symplectic eigenvalues of the matrix $\Lambda_{PT} {\pmb
    \sigma} \Lambda_{PT}$ :
     \beq
     \eta^{\pm} = \frac{1}{\sqrt{2}}
     \left[ \Sigma({\pmb    \sigma}) \pm (\Sigma({\pmb    \sigma})^2
    - 4 det ({\pmb    \sigma}) )^{1/2} \right]
     \eeq
     with $\Sigma({\pmb    \sigma})  = det ({\bf A} )+ det ({\bf B
    }) - 2 det ({\bf C} )$.
     The logarithmic negativity 
     is an entanglement monotone. It is an upper bound on the
    distillable entanglement and a lower bound on the entanglement of
    formation. It measures the entanglement in units of entanglement bits (ebits), where one ebit is the entanglement present in a singlet state.
     Positivity of  $E_N >0$ is a necessary condition for entanglement. In the particular case of
    Gaussian states it is both a necessary and sufficient condition.

     \section{Correlations between two uniformly accelerated
    oscillators}\label{correlations}    
     To compute the elements of the covariance matrix first we have
    to    quantize the quantum field in Rindler coordinates and evaluate
    several     Minkowskian vacuum expectation values. We refer to
    \cite{Unruh76,GO} for detailed discussions of how to carry out these
    calculations, and summarize here very briefly the main points.
     \par          
     To lighten the formulas     we shall chose our length unit so
    that :    \beq    a=1\qquad\qquad .    \eeq    The quantum field
    being    massless, it decomposes into left ($\leftarrow$) and
    right     ($\rightarrow$) modes:    \beq
    \pmb{\Phi}(U,V)=\mathop{\Phi}\limits_{{\leftarrow}}(U)+\mathop{\Phi}\limits_{{\rightarrow}}(V)
    \qquad . \eeq    These modes themselves split into modes defined on the
    left (L) and    right (R) Rindler quadrants. For example :    \beq
    \mathop{\Phi}\limits_{{\leftarrow}}(U)={{\mathop{\Phi}\limits_{{\leftarrow}}}}\strut_{R}(u_{R})
    + {{\mathop{\Phi}\limits_{{\leftarrow}}}}\strut_{L}(u_{L})    \qquad .\eeq   
     We can decompose these operators in terms of Rindler modes, and
    Rindler creation and destruction operators $b^\dagger_{  u(v)_{R(L)}
    } (\lambda)$, $b_{  u(v)_{R(L)} } (\lambda)$:
     \beqa
    {{\mathop{\Phi}\limits_{{\leftarrow}}}}\strut_{R}(u)&=&\int_{0}^\infty\left(
    e^{-i\lambda u}b_{{u_{R}} } (\lambda)+e^{i\lambda
    u}b^\dagger_{{u_{R}} } (\lambda)\right) \frac 1
    {\sqrt{4\,\pi\,\lambda}}d\lambda \qquad  ,\\
    {{\mathop{\Phi}\limits_{{\leftarrow}}}}\strut_{L}(u)&=&\int_{0}^\infty\left(
    e^{i\lambda u}b_{{u_{L}} } (\lambda)+e^{-i\lambda
    u}b^\dagger_{{u_{L}} } (\lambda)\right) \frac 1
    {\sqrt{4\,\pi\,\lambda}}d\lambda \qquad .  \eeqa    We emphasize the sign
    change in the arguments of the exponentials when    we pass from the
    left to the right Rindler quadrant. It is the reflect    of the
    opposite $u$-time orientation in these quadrants. 
     \\
     We can also decompose the field in terms of Minkowski modes and
    Minkowski creation and destruction operators $b^\dagger_{  U(V) }
    (w)$, $b_{  U(V) } (w)$:
     \beq
    {\mathop{\Phi}\limits_{{\leftarrow}}}(U)=\int_{0}^\infty\left(
    e^{-i\, w\, U}b_{{U} } (w)+e^{+i\, w\,    U}b^\dagger_{{U} }
    (w)\right) \frac 1    {\sqrt{4\,\pi\,w}}dw    \qquad  .\eeq 
     The link    between these two decompositions    is provided by
    standard Bogoljubov transformations. Using the Bogoljubov
    transformations one shows that the Minkowski vacuum is perceived by
    the uniformly accelerated observer as being populated by a thermal
    bath of Rindler quanta at temperature $T_U= \frac{a}{2 \pi}$. 
     \par
     The Bogoljubov transformation also allows us to evaluate
    expectation values such as: 
       \beqa    \label{expvalRR}    \frac{1}{2}
    \langle\{\tilde\Phi_{R}(\lambda),\tilde\Phi_{R}(\lambda')\}\rangle&=&
    \langle\{
    {\mathop{\tilde\Phi}\limits_{\leftarrow}}\strut_{R}(\lambda),\mathop{\tilde\Phi}\limits_{\leftarrow}\strut_{R}(\lambda')\}\rangle
    =\frac 1{4\pi\lambda}\coth\lambda\pi\    
    \delta(\lambda+\lambda')\qquad  ,\\
    \label{expvalRL}
    \frac{1}{2}\langle\{\tilde\Phi_{R}(\lambda),\tilde\Phi_{L}(\lambda')\}\rangle&=&
    \langle\{
    {\mathop{\tilde\Phi}\limits_{\leftarrow}}\strut_{R}(\lambda),\mathop{\tilde\Phi}\limits_{\leftarrow}\strut_{L}(\lambda')\}\rangle
    =\frac 1{4\pi\lambda}\frac 
    1{\sinh\lambda\pi}\delta(\lambda+\lambda')\qquad  .
    \eeqa   
     These expressions are then used            
     to evaluate the correlation matrix between position and
    momentum variables expressed as in eqs.    (\ref{qR} ,\ref{pL}) : 
     \beqa  
     \label{qqRR}
     qq_{RR} \equiv  \langle
    q_{R}(\tau_{R})q_{R}(\tau_{R})\rangle&=&\frac{4\,\gamma\,      \omega
    }{m}\int_{-\infty}^{+\infty}    \frac{\lambda^{2}
    }{(\lambda-\bar\Omega_{+})(\lambda-\bar\Omega_{-})
    (\lambda-\Omega_{+})(\lambda-\Omega_{-})}\frac {\coth \lambda
    \pi}{4\,\lambda \,\pi}d\lambda\\        \eeqa    where we have
    introduced the notations:    \beq    \gamma=\frac{\epsilon^{2}
    }{4\,m\, \omega}\qquad,\qquad    \Omega_{\pm}= \omega(\
    i\gamma\pm\sqrt{1-\gamma^{2} })\qquad,\qquad    \bar\Omega_{\pm}=
    \omega(-i\gamma\pm\sqrt{1-\gamma^{2} })\qquad .    \eeq     
     As shown in the appendix, the integral over $\lambda$ can be
    carried out, giving a closed form for $qq_{RR}$. 
     \\
     Similarly the other correlators can be expressed as

     \beqa    \label{qqRL} 
     &&qq_{RL}(T) \equiv \langle
    q_{R}(\tau_{R})q_{L}(\tau_{L})\rangle=-\frac{4\,\gamma\,
    \omega}{m}\int_{-\infty}^{+\infty}    \frac{\lambda\, e^{-i\lambda
    T}}{(\lambda-\bar\Omega_{+})^{2} (\lambda-\bar\Omega_{-})^{2}
    }\frac {1}{4\,\pi\,\sinh \lambda     \pi}d\lambda 
    \qquad , \\     
     &&\label{qpRR}
     \frac 12 \langle \{q_{R}(\tau_{R}),p_{R}(\tau_{R})\}\rangle= i
    \, 4\,\gamma \, \omega^{2}  \int_{-\infty}^{+\infty}
    \frac{1}{(\lambda-\bar\Omega_{+})(\lambda-\bar\Omega_{-})
    (\lambda-\Omega_{+})(\lambda-\Omega_{-})}\frac {\coth \lambda
    \pi}{4\,\pi}d\lambda =0\qquad  ,\\ 
     && \label{qpRL}
     qp_{RL}(T) \equiv \frac 12 \langle\{
    q_{L}(\tau_{L})p_{R}(\tau_{R})\}\rangle=  i      \, 4\,\gamma \,
    \omega^{3}  \int_{-\infty}^{+\infty}    \frac{e^{-i\lambda     T}
    }{(\lambda-\bar\Omega_{+})^{2} (\lambda-\bar\Omega_{-})^{2}}\frac
    {1}{4\,\pi\,\sinh \lambda     \pi}d\lambda  \qquad .  \eeqa    
     In the last two expressions there is a pole at $\lambda =0$. We
    resolve the ambiguity in the resulting integrals by integrating in
    the sense of a principal part, thereby obtaining $0$ for the first
    integral and a closed form given in the appendix for $qp_{RL}$.   
     \par
     The computation of the momentum correlators are more delicate
    because they diverge due to the of a double pole at $\lambda =0$.
    We therefore introduce the infinite constant  
    \beq
    K=\frac{\gamma\,m\,  \omega }{\pi ^2}\int    
    \lambda^{-2}d\lambda\qquad  ,
    \eeq
    and so obtain: 
     \beqa  
     \label{ppRR}  K + pp_{RR} &\equiv&
     \langle    p_{R}(\tau_{R})p_{R}(\tau_{R})\rangle=4\,\gamma\,m\,
    \omega^{5}\int_{-\infty}^{+\infty}    \frac{1
    }{(\lambda-\bar\Omega_{+})(\lambda-\bar\Omega_{-})
    (\lambda-\Omega_{+})(\lambda-\Omega_{-})}\frac {\coth \lambda
    \pi}{4\,\lambda \,\pi}d\lambda \qquad ,
     \\  
     \label{ppRL}
     K + pp_{RL}(T) &\equiv&
     \frac 12\langle \{
    p_{R}(\tau_{R}),p_{L}(\tau_{L})\}\rangle=4\,\gamma\,m\,
    \omega^{5}\int_{-\infty}^{+\infty}    \frac{e^{-i\lambda T}
    }{(\lambda-\bar\Omega_{+})^{2} (\lambda-\bar\Omega_{-})^{2}
    }\frac {1}{4\,\lambda \,\pi\,\sinh \lambda     \pi}d\lambda\qquad 
    ,
     \eeqa      
     where $pp_{RR}$ and $pp_{RL}(T)$ are finite quantities whose
     closed forms are given in the appendix.
     \par
     We may  therefore  write the correlation matrix    as 
     \beq  \label{sigma}  {\pmb \sigma}=\left( \ba{cccc}    qq_{RR}&
    0 &qq_{RL}(T)& qp_{RL}(T)\\    0&K+pp_{RR}&qp_{RL}(T)&K+pp_{RL}(T)\\
    qq_{RL}(T)&qp_{RL}(T)&qq_{RR}&0\\
    qp_{RL}(T)&K+pp_{RL}(T)&0&K+pp_{RR}    \ea \right)\qquad .    \eeq    
     \par
     In this expression, except for $K$ which is infinite,
    all the other terms are finite functions, depending only of the dimensionless
    parameters $\omega$, $\gamma$, $T$, which can be
    interpreted as follows. The parameter $\omega$ is the ratio between
    the transition frequency of the oscillator and the acceleration $a$:
    $\omega \equiv \frac \omega a = \frac \omega {2 \pi T_U}$ where $T_U$
    is the Unruh temperature. When $\omega > 2 \pi$ the probability that
    the oscillator is excited will be exponentially small. We therefore
    expect any entanglement between the oscillators to disapear for 
    large values of $\omega$ (since entanglement requires 
    superpositions between several states). The parameter $\gamma \equiv
    \frac{\epsilon^{2} }{4\,m\, \omega}$ is the ratio between the line
    width $\Gamma
    =\frac{\epsilon^{2} }{4\,m}$ (the inverse lifetime) of the first excited state  of the oscillator and its transition
    frequency $\omega$. When $\gamma >1$ the oscillator is strongly
    coupled to the field, whereas when $\gamma<1$ the oscillator is
    weakly coupled to the field. This translates in the Heisenberg
    equations of motion into the difference between the free solution
    being over damped or oscillating as it decays. In what follows we
    shall only consider the  regime $\gamma <1$. Finally $T\equiv a
    (\tau_R - \tau_L)$ is the difference of the Rindler times along the
    two trajectories, in units of the inverse acceleration. It measures
    the Lorentz invariant distance of points on the two trajectories: $\Delta 
    s^{2}=4\,a^{-2}\cosh a\,T/2$.
     \par
     We have checked numerically that the correlation matrix obeys
    the positivity constraint eq. (\ref{pos}). Indeed when inserting eq.
    (\ref{sigma}) into eq. (\ref{pos}) we find that the resulting matrix
    has one infinite positive eigenvalue, and three finite eigenvalues
    which we found to be positive using the procedure outlined in the
    appendix. 
     \par
     Similarly we can consider the condition of positive partial
    transpose eq. (\ref{condent}). Once more, inserting eq. (\ref{sigma})
    into eq. (\ref{condent}) we find that the resulting matrix has one
    infinite positive eigenvalue, and three finite eigenvalues. When one
    of these eigenvalues becomes negative the state is entangled. We have
    also computed the logarithmic negativity $E_N$  which quantifies the
    degree of entanglement present in the system. We find that
    logarithmic negativity is always finite and independent of $K$ 
    (see eq.(\ref{etaexp})).
     \begin{figure}[t]
     \includegraphics[scale=0.6]{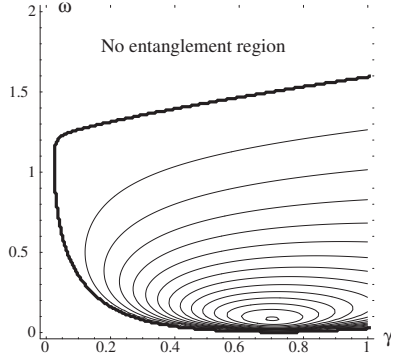}
     \caption{
     Maximum entanglement between the oscillators, as a function of
    the dimensionless parameters $\gamma$ and $\omega$. The outer line is
    the frontier of the region where the oscillators get entangled. The
    contour lines correspond to increases in the logarithmic negativity
    $E_N$ in steps of $0.1$. The maximum entanglement occurs for $\gamma
    \simeq 0.703$, $\omega \simeq 0.0845$ when $E_N = 1.406$.}
     \label{fig:ConPlot}
     \end{figure}
     \par
     Thus, even though the fluctuations of the oscillators coupled to
    the field are infinite, since the momentum correlators are infinite,
    the model is well defined. In particular the quantity we are
    interested in, the entanglement between the two oscillators, is
    always finite.  
     \par
     We have computed the entanglement between the two oscillators as
    a function of $T$ for different values of $\omega$ and $\gamma < 1$.
    We find that there are only specific pairs of values $(\gamma ,
    \omega)$ for which the detectors become entangled. In Fig.
    \ref{fig:ConPlot} we have plotted the pairs of values of $(\gamma ,
    \omega)$ for which the detector gets entangled. Note that our
    numerical analysis indicates that the region where entanglement
    occurs does not touch the axes $\gamma = 0$ and $\omega=0$. This is
    interesting since these axes correspond to the domain of validity of perturbation
    theory. Indeed the perturbative limit should arise when $\epsilon\to 0$, $m$
    fixed which corresponds to $\gamma \omega = \epsilon^2/m \to 0$. Thus the
    entanglement between two uniformly accelerated oscillators in
    opposite Rindler quadrants is a non perturbative phenomena.
     \par
\begin{figure}[t]
     \includegraphics[scale=0.6]{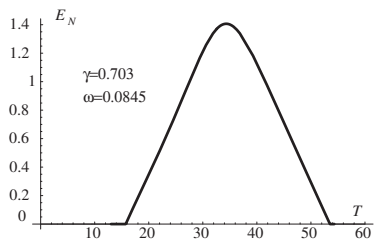}
     \caption{
     Evolution of the logarithmic negativity $E_N$ between the two
    oscillators as a function of $T = \tau_R - \tau_L$ for the values
    $\gamma = 0.703$, $\omega = 0.0845$ for which it reaches its
    maximal value of $1.406$. In these units the period
    of the oscillator is $2 \pi /\omega \simeq 74$, and the
    ''lifetime" of the first excited state of the oscillator is
    $\Gamma^{-1} = (\gamma \omega)^{-1} = 17$. (We put quotes because the
    concept of lifetime is not well defined when it is smaller than the
    oscillation period, although it serves as a useful guiding concept
    for understanding what are the different time scales involved). The oscillator is thus entangled for less than one oscillation period, and during this time it exchanges several quanta with the Unruh heat bath.
    Note also that in this case the Boltzman factor $e^{-2 \pi \omega}\simeq 0.6$ is of
    order 1
     which implies that several of the oscillator levels have
    significant probability of being populated -this is necessary since $E_N>1$ requires a system with dimension greater than 2-.
    The
    shapes of the curves for other values of $\gamma$ and $\omega$ 
    are
    similar, except that the maximum entanglements reached are smaller, and
    that the times at which these maxima are reached are different.}
     \label{fig:Timdep}
     \end{figure}

     In Fig. \ref{fig:ConPlot} we have also plotted the degree of
    entanglement between the two oscillators as a function of $(\gamma ,
    \omega)$. We see that the entanglement is maximum for $\gamma \simeq
    0.703$, $\omega \simeq 0.0845$ whereupon the logarithmic negativity
    reaches the value $E_N = 1.406$, see 
    Fig. \ref{fig:Timdep} for further discussion of this case. 
     \par 
     We have also computed how the entanglement between the two
    oscillators evolves as a function of $T$.  For all values of the
    parameters, we find that the entanglement only appears when $T>0$.
    That is the entanglement only gets established after a 
    configuration of
    closest approach ($T=\tau_{R}-\tau_{L}=0$) has been realized. The entanglement then increases, reaches a
    maximum,  decreases and goes zero at a finite value of $T$. We
    can understand this as follows. The oscillators emit and absorb quanta
    that are packets localised in frequency and time (for instance if $\gamma<<1$ then  $\Delta \lambda
    \simeq \epsilon^2/2m$ and $\Delta \tau \simeq
    (\epsilon^2/2m)^{-1}$). These quanta are only correlated around 
    configurations of
    closest approach $T\simeq 0$. It takes some time to establish
    correlations between the two detectors, which is why the entanglement
    only appears for $T>0$. At late times the quanta exchanged between
    the two oscillators are no longer correlated. The entanglement is
    gradually erased and finally disappears.  As illustration in Fig.
    \ref{fig:Timdep} we have plotted how the entanglement between the two
    detectors evolves as a function of $T$ for the value of $\gamma$ and
    $\omega$ for which entanglement is maximal
     \par
     Note that because of boost invariance there is in fact an
    infinite set of pairs of locations for which the oscillators are
    entangled.  Indeed if we look at one oscillator at a specific 
    value of its proper time $\tau_{L(R)}=\tau^\star$, it is only 
    entangled with the other for a finite interval of proper time 
    starting slightly after $\tau_{R(L)}=\tau^\star$. This is depicted schematically in Fig. \ref{fig:Traject}.
     \par
     In summary we have studied the entanglement between two
    uniformly accelerated oscillators in 1+1 dimensional Minkowsky space
    time coupled to a massless scalar field. This model is exactly
    soluble. It allows us to study the case where both detectors are in
    thermal equilibrium with the field. It also allows us to study the
    case where the detectors are strongly coupled to the field. We find
    that there are some choices of parameters and of positions along the
    trajectories for which the two detectors get entangled. The maximum
    entanglement we find is slightly larger than $1.4$ entanglement 
    bits. 
     
     \acknowledgments We thank the Fonds National de la Recherche
    Scientifique (FNRS) and its associated fund (FRFC) for financial support. S.M.
    also acknowledges financial support from the EU project FP6-511004
    COVAQIAL and integrated project QAP 015848.

     \section{Appendix: Details of Calculations}
     
     We group here some of the calculations that are behind our main
    results.
     \par
     \subsection{Explicit expressions for the correlation matrix
    elements}
     Here we give explicit expressions for the correlators, which are
    expressed in terms of integrals in the main text. Specifically, upon
    carrying out the integrals in eqs.
    (\ref{qqRR},\ref{qqRL},\ref{qpRL},\ref{ppRR},\ref{ppRL}), one finds
     \beqa      
     \langle    q_{R}(\tau_{R})q_{R}(\tau_{R})\rangle&=&
     \frac{\sinh \left(2 \pi  \sqrt{1-\gamma ^2} \omega \right)}{2
    \sqrt{1-\gamma ^2} \omega     \left(\cosh \left(2 \pi
    \sqrt{1-\gamma ^2} \omega \right)-\cos (2 \pi  \gamma  \omega
    )\right)}    -\frac{2 \gamma  \omega }{\pi }\sum_{n=1}^\infty
    \frac{n}{\left(n^2+\omega ^2\right)^2-4 n^2 \gamma ^2 \omega 
    ^2}\qquad ,
    \eeqa    
     \beqa    &&\langle   q_{R}(\tau_{R})q_{L}(\tau_{L})\rangle
     =\frac{\gamma\,  \theta (T)\,e^{-T \gamma   \omega  }}{
    \left(1-\gamma ^2\right)^{3/2}  \omega }     \Re\left[ \frac{e^{i T
    \sqrt{1-\gamma ^2}  \omega } \left( (1-\gamma ^2+i \gamma
    \sqrt{1-\gamma ^2} ) \omega  (T  +i \pi \,\coth \left[\pi   \omega
    (\sqrt{1-\gamma ^2}+i \gamma )\right])-\gamma \right)}{\sinh
    \left[\pi  (\sqrt{1-\gamma ^2}+i \gamma )  \omega \right]}\right]
    \nonumber\\
     && +\frac{2 \gamma   \omega  }{\pi }\left(
    \theta[T]\sum_{n=1}^\infty \frac{(-1)^n n e^{-n T}}    {\left(n^2-2
    \gamma   \omega   n+ \omega  ^2\right)^2}    +\theta[-T]
    \sum_{n=1}^\infty \frac{(-1)^n n e^{n T}}    {\left(n^2+2 \gamma
    \omega   n+ \omega      ^2\right)^2}\right) \qquad ,     \eeqa    
     \beqa       \frac 12 \langle\{
    &&q_{L}(\tau_{L})p_{R}(\tau_{R})\}\rangle
     =   
     -\frac{\gamma\,  \theta (T)\,e^{-T \gamma   \omega  }}{2
    \left(1-\gamma ^2\right)^{3/2} }     \Re\left[ \frac{e^{i T
    \sqrt{1-\gamma ^2}  \omega } \left( 1 +   \sqrt{1-\gamma ^2}  \omega
    ( \pi \,\coth \left[\pi   \omega  (\sqrt{1-\gamma ^2}+i \gamma
    )\right]-i T) \right)}{\sinh \left[\pi  (\sqrt{1-\gamma ^2}+i \gamma
    )  \omega \right]}\right] \\    
     &&\quad\quad\quad 
     +\frac{2 \gamma  \omega ^3}{\pi }\left[\theta[T]\left(
    \sum_{n=1}^\infty \frac{(-1)^n e^{-n T}}    {\left(n^2-2 \gamma
    \omega  n+\omega ^2\right)^2}+ \frac{1}{2     \omega
    ^4}\right)-\theta[-T]\left( \sum_{n=1}^\infty \frac{(-1)^n e^{n T}}
    {\left(n^2+2 \gamma  \omega  n+\omega ^2\right)^2}+ \frac{1}{2
    \omega ^4}\right)\right]  \quad ,
     \nonumber   \eeqa    
     \beqa    \langle    p_{R}(\tau_{R})p_{R}(\tau_{R})\rangle
     &=& K+ \frac{\omega  \left(\left(1-2 \gamma ^2 \right) \sinh
    \left(2 \pi     \sqrt{1-\gamma ^2} \omega \right)-2 \gamma
    \sqrt{1-\gamma ^2} \sin (2 \pi  \gamma  \omega )\right)}    {2
    \sqrt{1-\gamma    ^2}\left(\cosh \left(2 \pi   \sqrt{1-\gamma ^2}
    \omega \right)-\cos (2 \pi  \gamma  \omega )\right)}\nonumber	 \\
     & &+\frac{2 \gamma  \omega ^5}{\pi }\sum_{n=1}^\infty
    \frac{1}{n \left(\left(n^2+\omega ^2\right)^2-4 n^2 \gamma ^2 \omega
    ^2\right)}   \qquad , \eeqa   
     \beqa   && \frac 12\langle \{
    p_{R}(\tau_{R}),p_{L}(\tau_{L})\}\rangle
     =K+ \frac{\gamma\,  \theta (T)\,e^{-T \gamma   \omega
    }}{\left(1-\gamma ^2\right)^{3/2} }    \  \Re\left[e^{i T
    \sqrt{1-\gamma ^2}  \omega } \left( \frac{ 3\,\gamma -2 i \,\gamma
    ^{3}+2\,  (1-\gamma ^2)^{3/2} }{\sinh \left[\pi  (\sqrt{1-\gamma
    ^2}+i \gamma )  \omega \right]}\right .\right.\\ \nonumber 
     &&\qquad\qquad \qquad \qquad\qquad \qquad \left . \left . +
    \frac{(1-\gamma ^{2} -\,i \,\gamma   \sqrt{1-\gamma ^2} )\omega
    \left(  T  +i \pi \,\coth \left[\pi   \omega  (\sqrt{1-\gamma ^2}+i
    \gamma )\right] \right)}{\sinh \left[\pi  (\sqrt{1-\gamma ^2}+i
    \gamma )  \omega \right]}\right)\right] \\ \nonumber    
     &&+\frac{\gamma  \omega ^5}{\pi }\left[ \theta
    [T]\left(2\sum_{n=1}^\infty \frac{(-1)^n e^{-n T}}{n \left(n^2-2
    \gamma  \omega  n+\omega ^2\right)^2}    +\frac{4 \gamma   -T \omega
    }{\omega ^5}\right)+\theta    [-T]\left(2\sum_{n=1}^\infty
    \frac{(-1)^n e^{n T}}{n \left(n^2+2 \gamma  \omega  n+\omega
    ^2\right)^2}    -\frac{4 \gamma   -T \omega }{\omega ^5}\right)
    \right]\qquad . \nonumber   \eeqa        
     \par
     \subsection{Positivity and Entanglement}

     Positivity of the Hilbert space inner product implies the
    positivity of the matrix ${\pmb \sigma}+i {\pmb \Omega}$, see eq.
    (\ref{pos}). We checked that this is     indeed the case as one of
    the eigenvalue of this matrix is    infinite $= K$, while the three
    other are given by the    eigenvalues of the matrix acting on the
    orthogonal space to the    eigenvector of this infinite eigenvalue:

     \beq    \left(\begin{array}{ccc} {qq_{RR}}- {qq_{RL}} &
    \frac{i}{2}- {qp_{RL}} & 0 \\    -\frac{i}{2}- {qp_{RL}} &
    {pp_{RR}}-\ {pp_{RL}} & 0 \\       0 & 0 & {qq_{RL}}+
    {qq_{RR}}\end{array}\right)    \eeq    The three eigenvalues of this
    matrix are easy to compute. Using the above expressions for the
    correlators, we have    checked numerically that they are positive,
    as expected.         
     \par
     The criterium to put into evidence entanglement for a Gaussian system    consists to show the occurence of negative eigenvalue in the    partially transposed of the previous correlation matrix, {\it    i.e.} the negativity of the matrix    \beq \label{V}    {\pmb \sigma^{PT}  + i \Omega}=\left( \ba{cccc}    qqRR&qpRR+\frac{i}{2}&qqRL&-qpRL\\    qpRR-\frac{i}{2}&K+ppRR&qpRL&-K-ppRL\\    qqRL&qpRL&qqRR&-qpRR+\frac{i}{2}\\    -qpRL&-K-ppRL&-qpRR-\frac{i}{2}&K+ppRR    \ea \right)\qquad .    \eeq  
     Here again we find that one of the eigenvalues of this matrix is
    infinite $=K$.    The computation of the three other eigenvalues is
    less obvious than in    the previous case. First we may perform a
    symplectic    transformation, using the matrix    
     \beq    \left(\begin{array}{cccc}     \frac{1}{\sqrt{2}} & 0 &
    -\frac{1}{\sqrt{2}} & 0 \\     0 & \frac{1}{\sqrt{2}} & 0 &
    -\frac{1}{\sqrt{2}} \\     \frac{1}{\sqrt{2}} & 0 &
    \frac{1}{\sqrt{2}} & 0 \\     0 & \frac{1}{\sqrt{2}} & 0 &
    \frac{1}{\sqrt{2}}   \end{array}   \right)    \eeq    to obtain the
    expression    \beq\label{SV}    \left( \begin{array}{cccc}
    {qqRR}-{qqRL} & \frac{i}{2} & 0 & -{qpRL} \\    -\frac{i}{2} & 2
    K+{ppRL}+{ppRR} & {qpRL} & 0 \\    0 & {qpRL} & {qqRL}+{qqRR} &
    \frac{i}{2} \\    -{qpRL} & 0 & -\frac{i}{2} & {ppRR}-{ppRL}
    \end{array} \right)    \eeq   
     from which it is easy to isolate the eigenspace attached to the
    infinite eigenvalue and its orthogonal subspace. But to obtain the
    remaining three  eigenvalues, i.e. the eigenvalues of the reduced
    matrix   
     \beq   {\bf V} =\left(\begin{array}{ccc} {qqRR}- {qqRL} & 0&   -
    {qpRL}  \\   0&  {qqRL}+{qqRR} & \frac i2 \\       -qpRL & -\frac i2
    & {ppRR}- {ppRL}\end{array}\right) \ ,   \eeq 
     we have    to use (in principle) the general    Cardan formula.
    We have performed such an analysis numerically. We have also used the
    criteria eq. (\ref{condent}) which in the present case reduces to
     \beq   \lim_{K\to\infty}K^{-1}{\bf W}=-2 \det[{\bf    V}]\qquad
    . \label{W/K}   \eeq   
     Finally the smallest symplectic eigenvalue of the partial transpose of
    $\pmb \sigma$ can be expressed as:   \beq   \label{etaexp}
    \eta_{-}=
    \sqrt{\frac    14+\frac{\det [{\bf V}]}{(qqRR-qqRL)}}  \ . \eeq    It
    is independent of $K$. 
     The expressions (\ref{W/K}) and (\ref{etaexp}) were used to
    compute numerically the results discussed in the main text.
         \end{document}